\documentclass[letterpaper, 10 pt, conference]{ieeeconf}  

\IEEEoverridecommandlockouts                              

\usepackage{amsmath}
\usepackage{amssymb}
\usepackage{tikz}
\usepackage{pgfplots}
\usepackage{graphicx}
\usepackage{caption}
\usepackage{subcaption}
\usepackage{xcolor}
\allowdisplaybreaks

\newcommand{\until}[2]{\, U_{[{#1},{#2}]} \, }
\newcommand{\rs}[1]{\rho^{#1}(\boldsymbol{x},t)}
\newcommand{\rss}[2]{\rho^{#1}(\boldsymbol{x},{#2})}

\newcommand{\re}[1]{{\color{black}#1}}
\newcommand{\ree}[1]{{\color{black}#1}}

\newtheorem{definition}{Definition} 
\newtheorem{theorem}{Theorem} 
\newtheorem{assumption}{Assumption} 
\newtheorem{problem}{Problem} 
\newtheorem{remark}{Remark}
\newtheorem{lemma}{Lemma}
\newtheorem{corollary}{Corollary}

\title{\LARGE \bf
Decentralized Control Barrier Functions for Coupled Multi-Agent Systems  under Signal Temporal Logic Tasks
}

\author{Lars Lindemann and Dimos V. Dimarogonas
\thanks{This work was supported in part by the Swedish Research Council (VR), the European Research Council (ERC), the Swedish Foundation for Strategic Research (SSF),  the EU H2020 Co4Robots project, the SRA ICT TNG project STaRT, and the Knut and Alice Wallenberg Foundation (KAW).}
\thanks{The authors are with the Department of Automatic Control, School of Electrical Engineering and Computer Science, KTH Royal Institute of Technology, 100 44 Stockholm, Sweden. {\tt\small llindem@kth.se (L. Lindemann), dimos@kth.se (D.V. Dimarogonas)}}%
}

\begin{document}

\maketitle
\thispagestyle{empty}
\pagestyle{empty}

\begin{abstract}
We study the problem of controlling multi-agent systems under a set of signal temporal logic tasks. Signal temporal logic is a formalism that is used to express time and space constraints for dynamical systems. Recent methods to solve the control synthesis problem for single-agent systems under signal temporal logic tasks are, however, subject to a high computational complexity. Methods for multi-agent systems scale at least linearly with the number of agents and induce even higher computational burdens. We propose a computationally-efficient control strategy to solve the multi-agent control synthesis problem that results in a robust satisfaction of a set of signal temporal logic tasks. In particular, a decentralized feedback control law is proposed that is based on time-varying control barrier functions. The obtained control law is discontinuous and formal guarantees are provided by nonsmooth analysis. Simulations show the efficacy of the presented method. 
\end{abstract}

\section{Introduction}
\label{sec:introduction}

Control of multi-agent systems deals with achieving tasks such as consensus \cite{ren2005consensus}, formation control \cite{tanner2003stable}, and connectivity maintenance \cite{zavlanos2008distributed}. More recently, ideas from formal verification have been used where the task is formulated as a temporal logic formula. Signal temporal logic (STL) \cite{maler2004monitoring} allows to formulate time and space constraints and is based on continuous-time signals. It is hence suited to impose continuous-time tasks on continous-time systems. STL also admits robust semantics \cite{donze2} that state how robustly an STL formula is satisfied.  For discrete-time single-agent systems, \cite{raman1} and \cite{lindemann2016robust} propose optimization-based methods where in particular \cite{raman1} is based on a computationally expensive mixed integer linear program; \cite{raman1} has been extended to discrete-time multi-agent systems in \cite{liu2017distributed} and is hence subject to similar computational burdens. These methods fail to provide continuous time guarantees. A first step in this direction, i.e., for continuous-time systems, was made in \cite{pant2018fly} and \cite{lindemann2018decentralized}; \cite{pant2018fly}, however, is limited by the need of  solving a non-convex optimization problem, while the STL fragment in \cite{lindemann2018decentralized} is limited; \cite{lindemann2018control} establishes a connection between the semantics of an STL task and time-varying control barrier functions and derives a computationally-efficient feedback control law for single-agent systems. Control barrier functions have first been proposed in \cite{wieland2007constructive} and guarantee the existence of a control law that renders a desired set forward invariant; \cite{ames2017control} presents control barrier functions tailord for the safe robot navigation, while \cite{wang2017safety} presents decentralized control barrier functions for safe multi-robot navigation. Nonsmooth and time-varying control barrier functions have been proposed in \cite{glotfelter2017nonsmooth} and \cite{xu2018constrained}, respectively; \cite{srinivasan2018control} uses finite time control barrier functions for single-agent systems under linear temporal logic tasks, hence not allowing explicit timing constraints. 

We propose a decentralized control barrier function-based feedback control law for continuous-time multi-agent systems under a set of STL tasks. This control law is discontinuous. Therefore, we first extend our previous work on time-varying control barrier functions for single-agent systems, proposed in \cite{lindemann2018control}, to systems with discontinuous control inputs. We provide a barrier condition that, if satisfied, results in a satisfaction of an associated STL task. This barrier condition induces a certain load that needs to be accomplished by the control input. For multi-agent systems, we propose to share this load among agents by means of a discontinuous load sharing function resulting in a decentralized and discontinuous control law. Notably, each agent is subject to unknown dynamical couplings with other agents as well as noise. In the last step, we propose explicit construction rules for the control barrier functions that account for the STL semantics of a fragment of STL tasks. The construction of this barrier function can conveniently be carried out offline and maximizes the robustness by which the STL task is satisfied. The proposed decentralized feedback control law, which is calculated online, is obtained by solving a computationally tractable convex quadratic program.

Sec. \ref{sec:backgound} presents the problem formulation, while our proposed problem solution is stated in Sec. \ref{sec:strategy}. Simulations are presented in Sec. \ref{sec:simulations} followed by conclusions in Sec. \ref{sec:conclusion}.


\section{Preliminaries and Problem Formulation}
\label{sec:backgound}

True and false are $\top$ and $\bot$, respectively. Let $\boldsymbol{0}$ be a vector of appropriate size containing only zeros.  An extended class $\mathcal{K}$ function $\alpha:\mathbb{R}\to\mathbb{R}_{\ge 0}$ is a continuous and strictly increasing function with $\alpha(0)=0$. The partial derivative of a function $\mathfrak{b}(\boldsymbol{x},t):\mathbb{R}^n\times\mathbb{R}_{\ge 0}\to\mathbb{R}$ evaluated at $(\boldsymbol{x}',t')$ is abbreviated by  $\frac{\partial \mathfrak{b}(\boldsymbol{x}',t')}{\partial \boldsymbol{x}}:=\frac{\partial \mathfrak{b}( \boldsymbol{x},t)}{\partial \boldsymbol{x}}\Bigr|_{\substack{\boldsymbol{x}=\boldsymbol{x}'\\t=t'}}$ and assumed to be a row vector. For two sets $S_1$ and $S_2$, let $S_1\oplus S_2:=\{\boldsymbol{s}_1+\boldsymbol{s}_2|\boldsymbol{s}_1\in S_1, \boldsymbol{s}_2\in S_2\}$ denote the Minkowski sum.

\subsection{Discontinuous Systems and Nonsmooth Analysis}
Assume the system $\dot{\boldsymbol{x}}=f(\boldsymbol{x},t)$ where $f:\mathbb{R}^n\times\mathbb{R}_{\ge 0}\to\mathbb{R}^n$ is locally bounded and measurable. We consider Filippov solutions \cite{paden1987calculus} and define the Filippov set-valued map as
\begin{align*}
F[f](\boldsymbol{x},t)&:=\overline{\text{co}}\{\lim_{i\to\infty}f(\boldsymbol{x}_i,t)|\boldsymbol{x}_i\to \boldsymbol{x},\boldsymbol{x}_i\notin N\cup N_f  \}
\end{align*}
where $\overline{\text{co}}$ denotes the convex closure; $N_f$ denotes the set of Lebesgue measure zero where $f(\boldsymbol{x},t)$ is discontinuous, while $N$ denotes an arbitrary set of Lebesgue measure zero. A Filippov solution of $\dot{\boldsymbol{x}}=f(\boldsymbol{x},t)$ is an absolutely continuous function $\boldsymbol{x}:[t_0,t_1]\to\mathbb{R}^n$ that satisfies $\dot{\boldsymbol{x}}(t)\in F[f](\boldsymbol{x},t)$ for almost all $t\in[t_0,t_1]$. Due to \cite[Prop.~3]{cortes2008discontinuous} it holds that there exists a Filippov solution to $\dot{\boldsymbol{x}}=f(\boldsymbol{x},t)$ if $f:\mathbb{R}^n\times\mathbb{R}_{\ge 0}\to\mathbb{R}^n$ is locally bounded and measurable. Consider a continuously differentiable function $\mathfrak{b}(\boldsymbol{x},t)$ so that Clarke's generalized gradient of $\mathfrak{b}(\boldsymbol{x},t)$ coincides with the gradient of $\mathfrak{b}(\boldsymbol{x},t)$ \cite[Prop.~6]{cortes2008discontinuous}, denoted by $\nabla \mathfrak{b}(\boldsymbol{x},t):=\begin{bmatrix}
{\frac{\partial \mathfrak{b}(\boldsymbol{x},t)}{\partial \boldsymbol{x}}}^T & {\frac{\partial \mathfrak{b}(\boldsymbol{x},t)}{\partial t}}
\end{bmatrix}^T$. The set-valued Lie derivative of $\mathfrak{b}(\boldsymbol{x},t)$ is then defined as
\begin{align*}
\mathcal{L}_{F[f]}\mathfrak{b}(\boldsymbol{x},t)&:={\{\nabla \mathfrak{b}(\boldsymbol{x},t)}^T\begin{bmatrix}{\boldsymbol{\zeta}}^T & 1\end{bmatrix}^T|\boldsymbol{\zeta}\in F[f](\boldsymbol{x},t)\}.
\end{align*}
According to \cite[Thm.~2.2]{shevitz1994lyapunov}, it holds that $\dot{\mathfrak{b}}(\boldsymbol{x}(t),t)\in\mathcal{L}_{F[f]}\mathfrak{b}(\boldsymbol{x}(t),t)$ for almost all $t\in[t_0,t_1]$. Let $\hat{\mathcal{L}}_{F[f]}\mathfrak{b}(\boldsymbol{x},t):=\big\{\frac{\partial \mathfrak{b}(\boldsymbol{x},t)}{\partial \boldsymbol{x}}\boldsymbol{\zeta}|\boldsymbol{\zeta}\in F[f](\boldsymbol{x},t)\big\}$, the set-valued Lie derivative is then equivalent to $
\mathcal{L}_{F[f]}\mathfrak{b}(\boldsymbol{x},t)=\hat{\mathcal{L}}_{F[f]}\mathfrak{b}(\boldsymbol{x},t)\oplus \big\{\frac{\partial \mathfrak{b}(\boldsymbol{x},t)}{\partial t}\big\}$.
\begin{lemma}\label{lemma_liederivate}
Consider  $\dot{\boldsymbol{x}}=f_1(\boldsymbol{x},t)+f_2(\boldsymbol{x},t)$ where $f_1:\mathbb{R}^n\times\mathbb{R}_{\ge 0}\to\mathbb{R}^n$ and $f_2:\mathbb{R}^n\times\mathbb{R}_{\ge 0}\to\mathbb{R}^n$ are locally bounded and measurable. It then holds that 
\begin{small}
\begin{align*}
\mathcal{L}_{F[f_1+f_2]}\mathfrak{b}(\boldsymbol{x},t)\subseteq \hat{\mathcal{L}}_{F[f_1]}\mathfrak{b}(\boldsymbol{x},t)\oplus \hat{\mathcal{L}}_{F[f_2]}\mathfrak{b}(\boldsymbol{x},t)\oplus \Big\{\frac{\partial \mathfrak{b}(\boldsymbol{x},t)}{\partial t}\Big\}.
\end{align*}
\end{small}\begin{proof}
Holds by definition of $\mathcal{L}_{F[f_1+f_2]}\mathfrak{b}(\boldsymbol{x},t)$ since {\begin{small}$F[f_1+f_2](\boldsymbol{x},t)\subseteq F[f_1](\boldsymbol{x},t)\oplus F[f_2](\boldsymbol{x},t)$\end{small}} \cite[Thm. 1]{paden1987calculus}.
\end{proof}
\end{lemma}

\subsection{Signal Temporal Logic (STL)}
Signal temporal logic \cite{maler2004monitoring} is based on predicates $\mu$ that are obtained after evaluation of a continuously differentiable predicate function $h:\mathbb{R}^n\to\mathbb{R}$ as $\mu:=
 \begin{cases} 
 \top \text{ if } h(\boldsymbol{\zeta})\ge 0\\
 \bot \text{ if } h(\boldsymbol{\zeta})< 0
 \end{cases}$ for $\boldsymbol{\zeta}\in\mathbb{R}^n$. The STL syntax is then given by
\begin{align*}
\phi \; ::= \; \top \; | \; \mu \; | \; \neg \phi \; | \; \phi' \wedge \phi'' \; | \; \phi'  \until{a}{b} \phi''\;,
\end{align*}
where $\phi'$, $\phi''$ are STL formulas with $a\le b<\infty$ for $\until{a}{b}$ (until operator). Also define $F_{[a,b]}\phi:=\top \until{a}{b} \phi$ (eventually operator) and $G_{[a,b]}\phi:=\neg F_{[a,b]}\neg \phi$ (always operator). Let $\boldsymbol{x}\models \phi$ denote if the signal $\boldsymbol{x}:\mathbb{R}_{\ge 0}\to\mathbb{R}^n$ satisfies  $\phi$. These STL semantics are defined in  \cite{maler2004monitoring}; $\phi$ is satisfiable if $\exists \boldsymbol{x}:\mathbb{R}_{\ge 0}\to\mathbb{R}^n$ such that $\boldsymbol{x}\models \phi$. Robust semantics determine how robustly a signal $\boldsymbol{x}$ satisfies the formula $\phi$ at time $t$. The robust semantics for STL \cite[Def. 3]{donze2} are defined by:
\begin{small}
\begin{align*}
\rs{\mu}& := h(\boldsymbol{x}(t)), \hspace{0.2cm} \rs{\neg\phi} := 	-\rs{\phi},\\
\rs{\phi' \wedge \phi''} &:= 	\min(\rs{\phi'},\rs{\phi''}),\\
\rs{\phi' \until{a}{b} \phi''} &:= \underset{\bar{t}\in [t+a,t+b]}{\max}  \min(\rss{\phi''}{\bar{t}}, \underset{\underline{t}\in[t,\bar{t}]}{\min}\rss{\phi'}{\underline{t}} ),  \\
\rs{F_{[a,b]} \phi} &:= \underset{\bar{t}\in[t+a,t+b]}{\max}\rss{\phi}{\bar{t}},\\
\rs{G_{[a,b]} \phi} &:= \underset{\bar{t}\in[t+a,t+b]}{\min}\rss{\phi}{\bar{t}}.
\end{align*}
\end{small}Furthermore, it holds that $\boldsymbol{x}\models \phi$ if $\rho^\phi(\boldsymbol{x},0)>0$. 

\subsection{Coupled Multi-Agent Systems}

Consider  $M$ agents modeled by an undirected graph $\mathcal{G}:=(\mathcal{V},\mathcal{E})$ where $\mathcal{V}:=\{1,\hdots,M\}$ while $\mathcal{E}\in \mathcal{V}\times \mathcal{V}$ indicates communication links. Let $\boldsymbol{x}_i\in\mathbb{R}^{n_i}$ and $\boldsymbol{u}_i\in \mathbb{R}^{m_i}$ be the state and input of agent $i$. Also let $\boldsymbol{x}:=\begin{bmatrix} {\boldsymbol{x}_1}^T & \hdots & {\boldsymbol{x}_M}^T\end{bmatrix}^T\in\mathbb{R}^n$ with $n:=n_1+\hdots+n_M$ and
\begin{align}\label{system_noise}
\dot{\boldsymbol{x}}_i&=f_i(\boldsymbol{x}_i,t)+g_i(\boldsymbol{x}_i,t)\boldsymbol{u}_i+c_i(\boldsymbol{x},t)
\end{align}
where $f_i:\mathbb{R}^{n_i}\times \mathbb{R}_{\ge 0}\to\mathbb{R}^{n_i}$, $g_i:\mathbb{R}^{n_i}\times \mathbb{R}_{\ge 0}\to\mathbb{R}^{n_i\times m_i}$, and $c_i:\mathbb{R}^{n}\times \mathbb{R}_{\ge 0}\to\mathbb{R}^{n_i}$ are sufficiently regular. By sufficiently regular, we mean locally Lipschitz continuous and measurable in the first and second argument, respectively; $c_i(\boldsymbol{x},t)$ may model \emph{given dynamical couplings} such as those induced by a mechanical connection between agents; $c_i(\boldsymbol{x},t)$ may also describe unmodeled dynamics or process noise. We assume that $c_i(\boldsymbol{x},t)$ is bounded, but otherwise unknown so that no knowledge of $\boldsymbol{x}$ and $c_i(\boldsymbol{x},t)$ is required by agent $i$ for the control design. In other words, there exists $C\ge 0$ such that $\|c_i(\boldsymbol{x},t)\|\le C$ for all $(\boldsymbol{x},t)\in\mathbb{R}^{n}\times\mathbb{R}_{\ge 0}$.
\begin{assumption}\label{ass1}
The function $g_i(\boldsymbol{x}_i,t)$ has full row rank for all $(\boldsymbol{x}_i,t)\in\mathbb{R}^{n_i}\times\mathbb{R}_{\ge 0}$.
\end{assumption}
\begin{remark}\label{rem:1}
Assumption \ref{ass1} implies $m_i\ge n_i$. Since $\boldsymbol{x}$ and $c_i(\boldsymbol{x},t)$ are not known by agent~$i$, the system \eqref{system_noise} is \emph{not} feedback equivalent to $\dot{\boldsymbol{x}}_i=\boldsymbol{u}_i$. Canceling $f_i(\boldsymbol{x}_i)$ may also induce high control inputs, while we derive a minimum norm controller in Section \ref{sec:controller}. Collision avoidance, consensus, formation control, or connectivity maintenance can be achieved through a secondary controller $f_i^\text{u}$. Let therefore $\mathcal{V}_i^\text{u}\subseteq\mathcal{V}$ be a set of agents that \emph{induce dynamical couplings}, and let  $\boldsymbol{x}_i^\text{u}:=\begin{bmatrix} {\boldsymbol{x}_{j_1}}^T & \hdots & {\boldsymbol{x}_{j_{|\mathcal{V}_i^\text{u}|}}}^T\end{bmatrix}^T$ and $n_i^\text{u}:=n_{j_1}+\hdots+n_{j_{|\mathcal{V}_i^\text{u}|}}$ for $j_1,\hdots,j_{|\mathcal{V}_i^\text{u}|}\in\mathcal{V}_i^\text{u}$. By using $\boldsymbol{u}_i:={g_i(\boldsymbol{x}_i,t)}^T(g_i(\boldsymbol{x}_i,t){g_i(\boldsymbol{x}_i,t)}^T)^{-1}f_i^\text{u}(\boldsymbol{x}_i^\text{u},t)+\boldsymbol{v}_i$   the dynamics $\dot{\boldsymbol{x}}_i=f_i(\boldsymbol{x}_i,t)+f_i^\text{u}(\boldsymbol{x}^\text{u},t)+g_i(\boldsymbol{x}_i,t)\boldsymbol{v}_i+c_i(\boldsymbol{x},t)$ resemble \eqref{system_noise} if $f_i^\text{u}:\mathbb{R}^{n_i^\text{u}}\times\mathbb{R}_{\ge 0}\to\mathbb{R}^{n_i}$ is sufficiently regular.
\end{remark}

\subsection{Problem Formulation}

In this paper, we consider the STL fragment
\begin{subequations}\label{eq:subclass}
\begin{align}
\psi \; &::= \; \top \; | \; \mu \; | \; \neg \mu \; | \; \psi' \wedge \psi'' \label{eq:psi_class}\\
\phi \; &::= \;  G_{[a,b]}\psi \; | \; F_{[a,b]} \psi \;|\; \psi'  \until{a}{b} \psi'' \; | \; \phi' \wedge \phi''\label{eq:phi_class}
\end{align}
\end{subequations}
where $\psi'$, $\psi''$ are formulas of class $\psi$ in \eqref{eq:psi_class}, whereas $\phi'$, $\phi''$ are formulas of class $\phi$ in \eqref{eq:phi_class}. 
Consider $K$ formulas $\phi_1,\hdots,\phi_K$ of the form \eqref{eq:phi_class} and let the satisfaction of $\phi_k$ for $k\in\{1,\hdots,K\}$ depend on the set of agents ${\mathcal{V}}_k\subseteq \mathcal{V}$.
\begin{assumption}\label{ass:com}
For each $\phi_k$ with $k\in\{1,\hdots,K\}$, it holds that $(i_1,i_2)\in\mathcal{E}$ for all $i_1,i_2\in{\mathcal{V}}_k$.
\end{assumption}
 
Assume further that the sets of agents ${\mathcal{V}}_1,\hdots, {\mathcal{V}}_K\in\mathcal{V}$ are disjoint, i.e., ${\mathcal{V}}_{k_1}\cap{\mathcal{V}}_{k_2}=\emptyset$ for all $k_1,k_2\in\{1,\hdots,K\}$ with $k_1\neq k_2$, and such that ${\mathcal{V}}_1\cup\hdots\cup{\mathcal{V}}_K=\mathcal{V}$.

\begin{problem}\label{prob1}
Consider $K$ formulas $\phi_k$ of the form \eqref{eq:phi_class}. Derive a control law $\boldsymbol{u}_i$ for each agent $i\in\mathcal{V}$ so that $0<r\le \rho^{\phi_1 \wedge \hdots \wedge \phi_K}(\boldsymbol{x},0)$ where $r$ is maximized.
\end{problem}

\section{Control Approach}
\label{sec:strategy}

Section \ref{background_bf} extends \cite{lindemann2018control} to single-agent systems with discontinuous inputs. Based on this, Section \ref{sec:controller} proposes a decentralized feedback control law for multi-agent systems. Sections \ref{background_bf} and \ref{sec:controller} assume that the control barrier function satisfies some conditions that account for $\phi_k$. These conditions are addressed in Section \ref{sec:construction_rules}.

\subsection{Control Barrier Functions for Single-Agent Systems}
\label{background_bf}

For now, assume a single-agent system ($M=1$) given by
\begin{align}\label{eq:simplified_dynamics}
\dot{x}=f(\boldsymbol{x},t)+g(\boldsymbol{x},t)\boldsymbol{u}(\boldsymbol{x},t)+c(\boldsymbol{x},t)
\end{align}
where  $\boldsymbol{u}(\boldsymbol{x},t)$ may be a discontinuous function, while $f:\mathbb{R}^n\times\mathbb{R}_{\ge 0}\to\mathbb{R}^n$, $g:\mathbb{R}^n\times\mathbb{R}_{\ge 0}\to\mathbb{R}^{n\times m}$, and $c:\mathbb{R}^n\times\mathbb{R}_{\ge 0}\to\mathbb{R}^n$ are sufficiently regular, and where again there exists $C\ge 0$ such that $\|c(\boldsymbol{x},t)\|\le C$ for all $(\boldsymbol{x},t)\in\mathbb{R}^n\times\mathbb{R}_{\ge 0}$. This section does not require Assumption \ref{ass1}. Assume that \eqref{eq:simplified_dynamics} is subject to a formula $\phi$ of the form \eqref{eq:phi_class}.  

Consider  a function $\mathfrak{b}:\mathbb{R}^n\times[s_j,s_{j+1})\to\mathbb{R}$ that is continuously differentiable on $\mathbb{R}^n\times(s_j,s_{j+1})$ and let $\boldsymbol{x}:[t_j,t_{j+1}]\to\mathbb{R}^n$ be a Filippov solution to \eqref{eq:simplified_dynamics} under $\boldsymbol{u}(\boldsymbol{x},t)$ and with $t_j=s_j$. We distinguish between $t_{j+1}$ and $s_{j+1}$ since we want to ensure closed-loop properties over $[s_j,s_{j+1}]$, while $\boldsymbol{x}(t)$ may only be defined for $t_{j+1}<s_{j+1}$.  Also, define $
\mathfrak{C}(t):=\{\boldsymbol{x}\in \mathbb{R}^n | \mathfrak{b}(\boldsymbol{x},t)\ge 0\}$ and consider an open set $\mathfrak{D}\in\mathbb{R}^n$ such that $\mathfrak{D}\supset\mathfrak{C}(t)$ or all $t\in[s_j,s_{j+1})$. 

\begin{definition}[Control Barrier Function]\label{cCBF}
The function $\mathfrak{b}:\mathbb{R}^n\times[s_j,s_{j+1})\to \mathbb{R}$ is a candidate control barrier function (cCBF) for $[s_j,s_{j+1})$ if, for each $\boldsymbol{\boldsymbol{x}}_{0}\in\mathfrak{C}(s_j)$, there exists an absolutely continuous function $\boldsymbol{\boldsymbol{x}}:[s_j,s_{j+1})\to\mathbb{R}^n$ with $\boldsymbol{\boldsymbol{x}}(s_j):=\boldsymbol{\boldsymbol{x}}_0$ such that $\boldsymbol{\boldsymbol{x}}(t)\in\mathfrak{C}(t)$ for all $t\in [s_j,s_{j+1})$. A cCBF  $\mathfrak{b}(\boldsymbol{x},t)$ for $[s_j,s_{j+1})$ is a valid control barrier function (vCBF) for $[s_j,s_{j+1})$ and for  \eqref{eq:simplified_dynamics} under  $\boldsymbol{u}(\boldsymbol{x},t)$ if, for each function $c:\mathbb{R}^n\times\mathbb{R}_{\ge 0}\to\mathbb{R}^n$ with $\|c(\boldsymbol{x},t)\|\le C$, $\boldsymbol{x}_0\in\mathfrak{C}(t_j)$ implies, for each Filippov solution $\boldsymbol{x}:[t_j,t_{j+1}]\to\mathbb{R}^n$ to \eqref{eq:simplified_dynamics} under $\boldsymbol{u}(\boldsymbol{x},t)$ with $t_j=s_j$ and $\boldsymbol{x}(t_j)=\boldsymbol{x}_0$, $\boldsymbol{x}(t)\in\mathfrak{C}(t)$ for all $t\in[t_j,\min(t_{j+1},s_{j+1}))$.
\end{definition}

\begin{theorem}\label{theorem:disc_barrier}
Assume that $\mathfrak{b}(\boldsymbol{x},t)$ is a cCBF for $[s_j,s_{j+1})$. If $\boldsymbol{u}(\boldsymbol{x},t)$ is locally bounded and measurable and there is an extended locally Lipschitz continuous class $\mathcal{K}$ function $\alpha$ s.t.
\begin{align}\label{eq:barrier_ineq}
\min \mathcal{L}_{F[f+g\boldsymbol{u}]}\mathfrak{b}(\boldsymbol{x},t) \ge -\alpha(\mathfrak{b}(\boldsymbol{x},t))+\big\|\frac{\partial \mathfrak{b}(\boldsymbol{x},t)}{\partial \boldsymbol{x}}\big\|C
\end{align} 
for all $(\boldsymbol{x},t)\in \mathfrak{D}\times (s_j,s_{j+1})$, then $\mathfrak{b}(\boldsymbol{x},t)$ is a vCBF for $[s_j,s_{j+1})$ and for \eqref{eq:simplified_dynamics} under $\boldsymbol{u}(\boldsymbol{x},t)$.

\begin{proof}
Note that $-\frac{\partial \mathfrak{b}(\boldsymbol{x},t)}{\partial \boldsymbol{x}}c(\boldsymbol{x},t)\le |{\frac{\partial \mathfrak{b}(\boldsymbol{x},t)}{\partial \boldsymbol{x}}}c(\boldsymbol{x},t)|\le \|{\frac{\partial \mathfrak{b}(\boldsymbol{x},t)}{\partial \boldsymbol{x}}}\|\|c(\boldsymbol{x},t)\|\le \|\frac{\partial \mathfrak{b}(\boldsymbol{x},t)}{\partial \boldsymbol{x}}\|C$ so that \eqref{eq:barrier_ineq} implies
\begin{align}\label{eq:theorem1_ineq}
\min \mathcal{L}_{F[f+g\boldsymbol{u}]}\mathfrak{b}(\boldsymbol{x},t) \oplus {\frac{\partial \mathfrak{b}(\boldsymbol{x},t)}{\partial \boldsymbol{x}}}c(\boldsymbol{x},t)\ge -\alpha(\mathfrak{b}(\boldsymbol{x},t))
\end{align} 
for each function $c:\mathbb{R}^n\times\mathbb{R}_{\ge 0}\to\mathbb{R}^n$ with $\|c(\boldsymbol{x},t)\|\le C$. Assume that $\boldsymbol{x}(t_j)\in\mathfrak{C}(t_j)$ and note that $\dot{\mathfrak{b}}(\boldsymbol{x}(t),t)\in\mathcal{L}_{F[f+g\boldsymbol{u}+c]}\mathfrak{b}(\boldsymbol{x}(t),t)$ for almost all $t\in(t_j,\min(t_{j+1},s_{j+1}))$ and consequently also for almost all $t\in[t_j,\min(t_{j+1},s_{j+1}))$. Due to \eqref{eq:theorem1_ineq} it holds that $\min \mathcal{L}_{F[f+g\boldsymbol{u}]}\mathfrak{b}(\boldsymbol{x}(t),t) + {\frac{\partial \mathfrak{b}(\boldsymbol{x}(t),t)}{\partial \boldsymbol{x}}}c(\boldsymbol{x}(t),t)\ge -\alpha(\mathfrak{b}(\boldsymbol{x}(t),t))$ and according to Lemma \ref{lemma_liederivate}, it holds that $\min \mathcal{L}_{F[f+g\boldsymbol{u}+c]}\mathfrak{b}(\boldsymbol{x}(t),t)\ge \min\{ \mathcal{L}_{F[f+g\boldsymbol{u}]}\mathfrak{b}(\boldsymbol{x}(t),t) \oplus \hat{\mathcal{L}}_{F[c]}\mathfrak{b}(\boldsymbol{x}(t),t)\}=\min \mathcal{L}_{F[f+g\boldsymbol{u}]}\mathfrak{b}(\boldsymbol{x}(t),t)\oplus{\frac{\partial \mathfrak{b}(\boldsymbol{x}(t),t)}{\partial \boldsymbol{x}}}c(\boldsymbol{x}(t),t)$ since $\hat{\mathcal{L}}_{F[c]}\mathfrak{b}(\boldsymbol{x}(t),t)=\{\frac{\partial \mathfrak{b}(\boldsymbol{x}(t),t)}{\partial \boldsymbol{x}}c(\boldsymbol{x}(t),t)\}$ (due to \cite[Thm.~1]{paden1987calculus} and since $\mathfrak{b}(\boldsymbol{x},t)$ is continuously differentiable). It then holds that $\dot{\mathfrak{b}}(\boldsymbol{x}(t),t)\ge \min \mathcal{L}_{F[f+g\boldsymbol{u}+c]}\mathfrak{b}(\boldsymbol{x}(t),t) \ge -\alpha(\mathfrak{b}(\boldsymbol{x}(t),t))$. By \cite[Lem.~2]{glotfelter2017nonsmooth} it follows that $\mathfrak{b}(\boldsymbol{x}(t),t)\ge 0$ for all $t\in[t_j,\min(t_{j+1},s_{j+1}))$.
\end{proof}
\end{theorem}

In \cite{lindemann2018control}, a connection between $\mathfrak{b}(\boldsymbol{x},t)$ and the STL semantics of $\phi$ was established. In particular, conditions on $\mathfrak{b}(\boldsymbol{x},t)$ are imposed so that $\mathfrak{b}(\boldsymbol{x}(t),t)\ge 0$, i.e., $\boldsymbol{x}(t)\in\mathfrak{C}(t)$, for all $t\ge 0$  implies $\boldsymbol{x}\models \phi$. We shortly recall the main idea of \cite{lindemann2018control}.   In order to use conjunctions in $\phi$, a smooth under-approximation of the min-operator is used. For a number of $\tilde{p}$ cCBF's $\mathfrak{b}_l(\boldsymbol{x},t)$, note that
$\min_{l\in\{1,\hdots,\tilde{p}\}}\mathfrak{b}_l(\boldsymbol{x},t)\approx -\frac{1}{\eta}\ln\big(\sum_{l=1}^{\tilde{p}}\exp(-\eta \mathfrak{b}_l(\boldsymbol{x},t))\big)$ with $\eta>0$ and where the accuracy of this approximation increases as $\eta$ increases. In fact, it holds that $\lim_{\eta\to\infty}-\frac{1}{\eta}\ln\big(\sum_{l=1}^{\tilde{p}} \exp(-\eta\mathfrak{b}_l(\boldsymbol{x},t))\big)=\min_{l\in\{1,\hdots,\tilde{p}\}} \mathfrak{b}_l(\boldsymbol{x},t)$. Regardless of the choice of $\eta$, we have
\begin{align}\label{eq:under_approx}
-\frac{1}{\eta}\ln\big(\sum_{l=1}^{\tilde{p}} \exp(-\eta\mathfrak{b}_l(\boldsymbol{x},t))\big)\le \min_{l\in\{1,\hdots,\tilde{p}\}} \mathfrak{b}_l(\boldsymbol{x},t) 
\end{align}
Now, $\mathfrak{b}(\boldsymbol{x},t)\ge 0$ implies  $\mathfrak{b}_l(\boldsymbol{x},t)\ge 0$ for each $l\in\{1,\hdots,\tilde{p}\}$, i.e., the conjunction operator can be encoded. The conditions imposed on $\mathfrak{b}(\boldsymbol{x},t)$ are summarized in three steps (Steps A, B, and C).  For negations on predicates $\neg \mu$ as in $\eqref{eq:psi_class}$, let the corresponding predicate function be $-h(\boldsymbol{x})$. Furthermore, let $h_1(\boldsymbol{x})$, $h_2(\boldsymbol{x})$, $h_3(\boldsymbol{x})$, and $h_4(\boldsymbol{x})$ correspond to the atomic propositions $\mu_1$, $\mu_2$, $\mu_3$, and $\mu_4$, respectively.  

\textbf{Step A)} Consider single temporal operators in \eqref{eq:phi_class} that \emph{do not} contain conjunctions, i.e., $G_{[a,b]}\mu_1$, $F_{[a,b]} \mu_1$, and $\mu_1  \until{a}{b} \mu_2$. For $G_{[a,b]}\mu_1$, select $\mathfrak{b}(\boldsymbol{x},t)$ so that $\mathfrak{b}(\boldsymbol{x},t^\prime)\le h_1(\boldsymbol{x})$ for all $t^\prime\in[a,b]$. For $F_{[a,b]} \mu_1$, select $\mathfrak{b}(\boldsymbol{x},t)$ so that $\mathfrak{b}(\boldsymbol{x},t^\prime)\le h_1(\boldsymbol{x})$ for some $t^\prime\in[a,b]$. For $\mu_1  \until{a}{b} \mu_2$, select $\mathfrak{b}(\boldsymbol{x},t):=-\frac{1}{\eta}\ln\big(\exp(-\eta\mathfrak{b}_1(\boldsymbol{x},t))+\exp(-\eta\mathfrak{b}_2(\boldsymbol{x},t))\big)$ so that $\mathfrak{b}_2(\boldsymbol{x},t^\prime)\le h_2(\boldsymbol{x})$ for some $t^\prime\in[a,b]$ and $\mathfrak{b}_1(\boldsymbol{x},t^{\prime\prime})\le h_1(\boldsymbol{x})$ for all $t^{\prime\prime}\in[a,t^\prime]$. \textbf{Step B)} Consider single temporal operators in \eqref{eq:phi_class} that \emph{do} contain conjunctions, i.e., $G_{[a,b]}\psi_1$, $F_{[a,b]} \psi_1$, and $\psi_1  \until{a}{b} \psi_2$ where $\psi_1$ and $\psi_2$ may contain a conjunction of predicates as in \eqref{eq:psi_class}. Assume, without loss of generality, $\psi_1:=\mu_1\wedge\mu_2$ and $\psi_2:=\mu_3\wedge\mu_4$. For $G_{[a,b]}\psi_1$, select $\mathfrak{b}(\boldsymbol{x},t):=-\frac{1}{\eta}\ln\big(\exp(-\eta\mathfrak{b}_1(\boldsymbol{x},t))+\exp(-\eta\mathfrak{b}_2(\boldsymbol{x},t))\big)$ so that $\mathfrak{b}_1(\boldsymbol{x},t^\prime)\le h_1(\boldsymbol{x})$ and $\mathfrak{b}_2(\boldsymbol{x},t^\prime)\le h_2(\boldsymbol{x})$ for all $t^\prime\in[a,b]$. For $F_{[a,b]}\psi_1$, select $\mathfrak{b}(\boldsymbol{x},t):=-\frac{1}{\eta}\ln\big(\exp(-\eta\mathfrak{b}_1(\boldsymbol{x},t))+\exp(-\eta\mathfrak{b}_2(\boldsymbol{x},t))\big)$ so that $\mathfrak{b}_1(\boldsymbol{x},t^\prime)\le h_1(\boldsymbol{x})$ and $\mathfrak{b}_2(\boldsymbol{x},t^\prime)\le h_2(\boldsymbol{x})$ for some $t^\prime\in[a,b]$. For $\psi_1  \until{a}{b} \psi_2$, select $\mathfrak{b}(\boldsymbol{x},t):=-\frac{1}{\eta}\ln\big(\sum_{l=1}^4\exp(-\eta\mathfrak{b}_l(\boldsymbol{x},t))\big)$ so that $\mathfrak{b}_3(\boldsymbol{x},t^\prime)\le h_3(\boldsymbol{x})$ and $\mathfrak{b}_4(\boldsymbol{x},t^\prime)\le h_4(\boldsymbol{x})$ for some $t^{\prime}\in[a,b]$ and $\mathfrak{b}_1(\boldsymbol{x},t^{\prime\prime})\le h_1(\boldsymbol{x})$ and $\mathfrak{b}_2(\boldsymbol{x},t^{\prime\prime})\le h_2(\boldsymbol{x})$ for all $t^{\prime\prime}\in[a,t^\prime]$. \textbf{Step C)} Consider conjunctions of temporal operators. For instance, consider $(G_{[a_1,b_1]}\psi_1) \wedge (F_{[a_2,b_2]} \psi_2) \wedge (\psi_3  \until{a_3}{b_3} \psi_4)$. Let $\mathfrak{b}(\boldsymbol{x},t):=-\frac{1}{\eta}\ln\big(\sum_{l=1}^3\exp(-\eta\mathfrak{b}_l(\boldsymbol{x},t))\big)$ where $\mathfrak{b}_1(\boldsymbol{x},t)$, $\mathfrak{b}_2(\boldsymbol{x},t)$, and $\mathfrak{b}_3(\boldsymbol{x},t)$ are associated with one temporal operator each and constructed as in Steps A and B.

We integrate $\mathfrak{o}_l:\mathbb{R}_{\ge 0}\to\{0,1\}$ into $\mathfrak{b}(\boldsymbol{x},t):=-\frac{1}{\eta}\ln\big(\sum_{l=1}^p\mathfrak{o}_l(t)\exp(-\eta\mathfrak{b}_l(\boldsymbol{x},t))\big)$ to reduce conservatism as in \cite{lindemann2018control}; $p$ is the total number of functions $\mathfrak{b}_l(\boldsymbol{x},t)$ obtained from Steps A, B, and C and each $\mathfrak{b}_l(\boldsymbol{x},t)$ corresponds to either an always, eventually, or until operator with a corresponding time interval $[a_l,b_l]$. To reduce conservatism, we remove single functions $\mathfrak{b}_l(\boldsymbol{x},t)$ from $\mathfrak{b}(\boldsymbol{x},t)$ when the corresponding always, eventually, or until operator is satisfied. For each temporal operator, the associated $\mathfrak{b}_l(\boldsymbol{x},t)$ is removed at $t=b_l$, i.e., $\mathfrak{o}_l(t)=1$ if $t<b_l$ and $\mathfrak{o}_l(t):=0$ if $t\ge b_l$.  The function $\mathfrak{b}(\boldsymbol{x},t)$ is now piecewise continuous in $t$. We denote the switching sequence by $\{s_0:=0,s_1,\hdots,s_q\}$ with $q\in\mathbb{N}$ as the total number of switches. This sequence is known due to knowledge of $[a_l,b_l]$. At time $t\ge s_j$ we have $s_{j+1}:= \text{argmin}_{b_l\in\{b_1,\hdots,b_p\}}\zeta(b_l,t)$ where $\zeta(b_l,t):=b_l-t$ if $b_l-t> 0$ and $\zeta(b_l,t):=\infty$ otherwise. To guarantee Filippov solutions $\boldsymbol{x}:[t_0,t_1]\to\mathbb{R}^n$ with $t_1\ge s_q$, we finally require that $\boldsymbol{x}(t)\in\mathfrak{C}(t)$ implies $\boldsymbol{x}(t)\in\mathfrak{D}'\subset\mathfrak{D}$ where $\mathfrak{D}'$ is some compact set. This requirement is not restrictive and can be achieved by adding a function $\mathfrak{b}_{p+1}(\boldsymbol{x},t)$ to $\mathfrak{b}(\boldsymbol{x},t):=-\frac{1}{\eta}\ln\big(\sum_{l=1}^{p+1}\mathfrak{o}_l(t)\exp(-\eta\mathfrak{b}_l(\boldsymbol{x},t))\big)$ where $\mathfrak{b}_{p+1}(\boldsymbol{x},t)=D-\|\boldsymbol{x}\|$ for a suitably selected $D\ge 0$, i.e., $D$ is such that $\mathfrak{D}':=\{\boldsymbol{x}\in\mathbb{R}^{n}|\|\boldsymbol{x}\|\le D\}\subset \mathfrak{D}$.

\begin{corollary}\label{corollary_phi_sat}
Consider $\phi$ of the form \eqref{eq:phi_class} and the system \eqref{eq:simplified_dynamics}. Let $\mathfrak{b}(\boldsymbol{x},t)$ satisfy the conditions in Steps A, B, and C for $\phi$ and be a cCBF for each time interval $[s_j,s_{j+1})$. If $\boldsymbol{u}(\boldsymbol{x},t)$ is locally bounded, measurable, and such that \eqref{eq:barrier_ineq} holds for all $(\boldsymbol{x},t)\in\mathfrak{D}\times(s_j,s_{j+1})$, then it follows that $\boldsymbol{x}\models \phi$ for each Filippov solution to \eqref{eq:simplified_dynamics} under $\boldsymbol{u}(\boldsymbol{x},t)$.

\begin{proof}
It holds that $\lim_{\tau\to s_j^-}\mathfrak{C}(\tau)\subseteq\mathfrak{C}(s_j)$ where $\lim_{\tau\to s_j^-}\mathfrak{C}(\tau)$ denotes the left-sided limit of  $\mathfrak{C}(t)$ at $t=s_j$. It is hence sufficient to ensure forward invariance of $\mathfrak{C}(t)$ for each $[s_j,s_{j+1})$ separately. There exist Filippov solutions $\boldsymbol{x}:[t_0,t_1]\to\mathbb{R}^n$ to \eqref{eq:simplified_dynamics} from each $\boldsymbol{x}(t_0)\in\mathfrak{C}(t_0)$. Due to Theorem~\ref{theorem:disc_barrier}, it follows that $\boldsymbol{x}(t)\in\mathfrak{C}(t)$ for all $t\in[t_0,\min(t_1,s_1))$; $\boldsymbol{x}(t)\in\mathfrak{C}(t)$ implies $\boldsymbol{x}(t)\in\mathfrak{D}'\subset\mathfrak{D}$, i.e., $\boldsymbol{x}(t)$ remains in the compact set $\mathfrak{D}'$, which implies $t_1\ge s_1$ by \ree{\cite[Ch.~2.7]{filippov2013differential}}. The same reasoning can be applied for the intervals $[s_1,s_2)$, $\hdots$, $[s_{q-1},s_q)$. It follows that  $\boldsymbol{x}\models\phi$.
\end{proof}
\end{corollary}


\subsection{Control Barrier Functions for Multi-Agent Systems}
\label{sec:controller}
Consider again $M$ agents subject to \eqref{system_noise} and $K$ formulas $\phi_k$ of the form \eqref{eq:phi_class}. For $j_1,\hdots,j_{|{\mathcal{V}}_k|}\in{\mathcal{V}}_k$, let $\bar{\boldsymbol{x}}_k:=\begin{bmatrix}
{\boldsymbol{x}_{j_1}}^T & \hdots & {\boldsymbol{x}_{j_{|{\mathcal{V}}_k|}}}^T
\end{bmatrix}^T\in\mathbb{R}^{\bar{n}_k}$ and $\bar{g}_k(\bar{\boldsymbol{x}}_k,t):=\text{diag}({g_{j_1}(\boldsymbol{x}_{j_1},t)}, \hdots ,  {g_{j_{|\mathcal{V}_k|}}(\boldsymbol{x}_{j_{|\mathcal{V}_k|}},t)})$ where $\bar{n}_k:=n_{j_1}+\hdots+n_{j_{|{\mathcal{V}}_k|}}$. Let $\mathfrak{b}^k(\bar{\boldsymbol{x}}_k,t)$ be the cCBF corresponding to $\phi_k$ and accounting for Steps A, B, and C. Define $\mathfrak{C}_k(t):=\{\bar{\boldsymbol{x}}_k\in\mathbb{R}^{\bar{n}_k} | \mathfrak{b}^k(\bar{\boldsymbol{x}}_k,t)\ge 0\}$ and let $\{s_0^k:=0,s_1^k,\hdots,s_{q_k}^k\}$ be the switching sequence corresponding to $\mathfrak{b}^k(\bar{\boldsymbol{x}}_k,t)$ as discussed in the previous section. We next consider cases where ${\frac{\partial \mathfrak{b}^k(\bar{\boldsymbol{x}}_k,t)}{\partial \bar{\boldsymbol{x}}_k}}\bar{g}_k(\bar{\boldsymbol{x}}_k,t)= {\boldsymbol{0}}^T$ for $(\bar{\boldsymbol{x}}_k,t)\in \mathfrak{D}_k\times (s_j^k,s^k_{j+1})$ where $\mathfrak{D}_k$ is an open and bounded set such that $\mathfrak{D}_k\supset \mathfrak{C}_k(t)$ for all $t\ge 0$. These occurences mean that $\mathfrak{b}^k(\bar{\boldsymbol{x}}_k,t)$, although possibly being a cCBF for $[s_j^k,s^k_{j+1})$, may not be a vCBF for $[s_j^k,s^k_{j+1})$ and for \eqref{system_noise} under any control law since \eqref{eq:theorem1_ineq} might fail to hold. Due to Assumption \ref{ass1}, it holds  that ${\frac{\partial \mathfrak{b}^k(\bar{\boldsymbol{x}}_k,t)}{\partial \bar{\boldsymbol{x}}_k}}\bar{g}_k(\bar{\boldsymbol{x}}_k,t)= {\boldsymbol{0}}^T$ if and only if $\frac{\partial \mathfrak{b}^k(\bar{\boldsymbol{x}}_k,t)}{\partial \bar{\boldsymbol{x}}_k}=\boldsymbol{0}$.

\begin{assumption}\label{ass3}
For an extended locally Lipschitz continuous class $\mathcal{K}$ function $\alpha_k$ and for $(\bar{\boldsymbol{x}}_k,t)\in\mathfrak{D}_k\times(s^k_j,s^k_{j+1})$ with $\frac{\partial \mathfrak{b}^k(\bar{\boldsymbol{x}}_k,t)}{\partial \bar{\boldsymbol{x}}_k}= \boldsymbol{0}$ it holds that $\frac{\partial \mathfrak{b}^k(\bar{\boldsymbol{x}}_k,t)}{\partial t}> -\alpha_k(\mathfrak{b}^k(\bar{\boldsymbol{x}}_k,t))$. 
\end{assumption}

\begin{theorem}\label{theorem1}
Consider a multi-agent system consisting of $M$ agents that are subject to the dynamics in \eqref{system_noise} satisfying Assumption \ref{ass1}  and $K$ formulas $\phi_k$ of the form \eqref{eq:phi_class} satisfying Assumption \ref{ass:com}. Assume further that each $\mathfrak{b}^k(\bar{\boldsymbol{x}}_k,t)$  accounts for the STL semantics of $\phi_k$ according to Steps A, B, and C, and that $\mathfrak{b}^k(\bar{\boldsymbol{x}}_k,t)$ is a cCBF for each time interval $[s^k_j,s^k_{j+1})$ satisfying Assumption \ref{ass3}. If each agent $i\in{\mathcal{V}}_k$  applies the control law $\boldsymbol{u}_i^*(\bar{\boldsymbol{x}}_k,t):=\boldsymbol{u}_i$  where $\boldsymbol{u}_i$ is given by
\begin{small}
\begin{subequations}\label{eq:qp_conv}
\begin{align}
&\hspace{-0.1cm}\min_{\boldsymbol{u}_i}\; \boldsymbol{u}_i^T\boldsymbol{u}_i\\
\begin{split}
\hspace{-0.1cm}\text{s.t. }&    \hspace{-0.1cm}\frac{\partial \mathfrak{b}^k(\bar{\boldsymbol{x}}_k,t)}{\partial \boldsymbol{x}_i}(f_i(\boldsymbol{x}_i,t)+g_i(\boldsymbol{x}_i,t)\boldsymbol{u}_i)\ge\|{\frac{\partial \mathfrak{b}^k(\bar{\boldsymbol{x}}_k,t)}{\partial \boldsymbol{x}_i}}\|\hat{n}C\\
 & \hspace{-0.1cm}-\mathfrak{N}_i(\bar{\boldsymbol{x}}_k,t)\big(\frac{\partial \mathfrak{b}^k(\bar{\boldsymbol{x}}_k,t)}{\partial t} +\alpha_k(\mathfrak{b}^k(\bar{\boldsymbol{x}}_k,t))\big),\label{eq:const_qp}
\end{split}
\end{align}
\end{subequations}
\end{small}where $\hat{n}:=\sqrt{\bar{n}_k\max(n_1,\hdots,n_M)}$ and where
\begin{small} 
\begin{align*}
\mathfrak{N}_i(\bar{\boldsymbol{x}}_k,t):=
\begin{cases}
\frac{\| \frac{\partial \mathfrak{b}^k(\bar{\boldsymbol{x}}_k,t)}{\partial \boldsymbol{x}_i}\|}{\sum_{d\in{\mathcal{V}}_k}\| \frac{\partial \mathfrak{b}^k(\bar{\boldsymbol{x}}_k,t)}{\partial \boldsymbol{x}_d}\|} &\hspace{-0.3cm}\text{if } \sum_{d\in{\mathcal{V}}_k}\| \frac{\partial \mathfrak{b}^k(\bar{\boldsymbol{x}}_k,t)}{\partial \boldsymbol{x}_d}\|\neq 0\\
1 &\hspace{-0.3cm}\text{otherwise,}
\end{cases}
\end{align*} 
\end{small}then $\bar{\boldsymbol{x}}_k\models\phi_k$ for each Filippov solution to \eqref{system_noise}. 

\begin{proof}
 We only provide a sketch of the proof due to space limitations. It can be shown that $\boldsymbol{u}_i^*(\bar{\boldsymbol{x}}_k,t)$ is locally bounded and measureable so that Filippov solutions exist. It can then be shown that \eqref{eq:const_qp} is always feasible and implies
\begin{small}
\begin{align}\label{eq:barrier_cond_phik}
\begin{split}
&\min \tilde{\mathcal{L}}_{F[\bar{f}_k+\bar{g}_k\bar{\boldsymbol{u}}_k^*]}\mathfrak{b}^k(\bar{\boldsymbol{x}}_k,t) \ge\\
&\hspace{1cm} -\alpha_k(\mathfrak{b}^k(\bar{\boldsymbol{x}}_k,t))+\big\|\frac{\partial \mathfrak{b}^k(\bar{\boldsymbol{x}}_k,t)}{\partial \bar{\boldsymbol{x}}_k}\big\|\sqrt{\bar{n}_k}C
\end{split}
\end{align}
\end{small}where $\bar{f}_k(\bar{\boldsymbol{x}}_k,t):=\begin{bmatrix}{f_{j_1}(\boldsymbol{x}_{j_1},t)}^T & \hdots & {f_{j_{|\mathcal{V}_k|}}(\boldsymbol{x}_{j_{|\mathcal{V}_k|}},t)}^T\end{bmatrix}^T$ and $\bar{\boldsymbol{u}}_k^*(\bar{\boldsymbol{x}}_k,t):=\begin{bmatrix}
{\boldsymbol{u}_{j_1}^*}^T & \hdots & {\boldsymbol{u}_{j_{|{\mathcal{V}}_k|}}^*}^T
\end{bmatrix}^T$ for $j_1,\hdots,j_{|{\mathcal{V}}_k|}\in{\mathcal{V}}_k$ so that Corollary \ref{corollary_phi_sat} can be used (note that $\sqrt{\bar{n}_k}C$ is needed here instead of only $C$).\end{proof}
\end{theorem}

The control law  $\boldsymbol{u}_i^*(\bar{\boldsymbol{x}}_k,t)$ is discontinuous at points $(\bar{\boldsymbol{x}}_k,t)$ that lie at the intersection of the closures of the regions defined by  the following three cases: case 1 where $\sum_{d\in{\mathcal{V}}_k}\|\frac{\partial \mathfrak{b}^k(\bar{\boldsymbol{x}}_k,t)}{\partial \boldsymbol{x}_j}\|= 0$, case 2 where $\sum_{d\in{\mathcal{V}}_k}\|\frac{\partial \mathfrak{b}^k(\bar{\boldsymbol{x}}_k,t)}{\partial \boldsymbol{x}_j}\|\neq 0$ and $\frac{\partial \mathfrak{b}^k(\bar{\boldsymbol{x}}_k,t)}{\partial \boldsymbol{x}_i}=\boldsymbol{0}$, and case 3 where $\sum_{d\in{\mathcal{V}}_k}\|\frac{\partial \mathfrak{b}^k(\bar{\boldsymbol{x}}_k,t)}{\partial \boldsymbol{x}_j}\|\neq 0$ and $\frac{\partial \mathfrak{b}^k(\bar{\boldsymbol{x}}_k,t)}{\partial \boldsymbol{x}_i}\neq\boldsymbol{0}$. The load sharing function $\mathfrak{N}_i(\boldsymbol{x},t)$  distributes the work needed to satisfy \eqref{eq:barrier_cond_phik} among agents; \eqref{eq:qp_conv} is a computationally tractable convex quadratic program. Computation of $\boldsymbol{u}_i^*(\bar{\boldsymbol{x}}_k,t)$ is decentralized and information does not need to be sent from local sensors to a central control unit and back to the actuators of each agent. Information only needs to be sent from local sensors to the local controllers.

\subsection{Explicit Control Barrier Function Construction}
\label{sec:construction_rules}

Since the contruction procedure is the same for each $\phi_k$, we omit the index $k$ for readability reasons. First, define the function $\gamma_l:\mathbb{R}_{\ge 0}\to \mathbb{R}$ as $\gamma_l(t):=(\gamma_{l,0}-\gamma_{l,\infty})\exp(-\mathfrak{l}_lt)+\gamma_{l,\infty}$ where $\gamma_{l,0},\gamma_{l,\infty}\in \mathbb{R}$ and $\mathfrak{l}_l\in\mathbb{R}_{\ge 0}$. The function $\gamma_l(t)$ is associated with $h_l(\boldsymbol{x})$, which in turn corresponds to  $\mu_l$. We require that $h^{\text{opt}}_l\ge 0$ where $h^{\text{opt}}_l:=\sup_{\boldsymbol{x}\in\mathbb{R}^n} h_l(\boldsymbol{x})$ ($\mu_l$ needs to be satisfiable) and consider to satisfy $\phi$ with robustness $r\in\mathbb{R}_{\ge 0}$. The construction of $\mathfrak{b}(\boldsymbol{x},t)$ is discussed in two steps (Steps 1 and 2) and is based on the conditions given in Steps A, B, and C in Section \ref{background_bf} leading to a function \re{$\mathfrak{b}(\boldsymbol{x},t):=-\frac{1}{\eta}\ln\big(\sum_{l=1}^p\mathfrak{o}_l(t)\exp(-\eta\mathfrak{b}_l(\boldsymbol{x},t))\big)$} where each $\mathfrak{b}_l(\boldsymbol{x},t)$ is associated with either $F_{[a_l,b_l]}\mu_l$ or $G_{[a_l,b_l]}\mu_l$. Recall  that each $\mu_{l_1}  \until{a_l}{b_l} \mu_{l_2}$ is encoded as $F_{[b_l,b_l]}\mu_{l_1}\wedge G_{[a_l,b_l]}\mu_{l_2}$ as  discussed in Steps A, B, and C.

\textbf{Step 1)}  Assume $p=1$ and consider $\phi:=G_{[a_l,b_l]}\mu_l$ and $\phi:=F_{[a_l,b_l]}\mu_l$. It is first required that $r\le h_l^\text{opt}$. Then, set
\begin{align}\label{eq:t_star}
t^*_l:=\begin{cases}
b_l &\text{if}\;\;\; F_{[a_l,b_l]} \mu_l\\
a_l &\text{if}\;\;\; G_{[a_l,b_l]} \mu_l,
\end{cases}
\end{align} 
which reflects the requirement that $\mu_l$ holds at least once between $[a_l,b_l]$ for $F_{[a_l,b_l]} \mu_l$ or at all times within $[a_l,b_l]$ for $G_{[a_l,b_l]} \mu_l$. If $t^*_l=0$, it is naturally required that $h_l(\boldsymbol{x}(0))\ge r$.  Let now $\mathfrak{b}_l(\boldsymbol{x},t):=-\gamma_l(t)+h_l(\boldsymbol{x})$  with parameters
\begin{subequations}\label{eq:gamma}
\begin{align}
\gamma_{l,0} &\in \begin{cases}
\big(-\infty,h_l(\boldsymbol{x}(0))\big) &\text{if}\;\;\; t^*_l>0\\
\big[r,h_l(\boldsymbol{x}(0))\big) &\text{otherwise}
\end{cases}\label{eq:gamma0}\\
\gamma_{l,\infty} &\in(\re{\max(r,\gamma_{l,0})},h^{\text{opt}}_l)\label{eq:gammainfty}\\
\mathfrak{l}_l & \in \begin{cases}
-\ln\big(\frac{r-\gamma_{l,\infty}}{\gamma_{l,0}-\gamma_{l,\infty}}\big)/t^*_l &\text{if}\;\;\; \gamma_{l,0}<r\\
0 &\text{otherwise}\label{eq:l}.
\end{cases}
\end{align}
\end{subequations}
Note that by the choice of $\gamma_{l,0}$, it holds that $\mathfrak{b}_l(\boldsymbol{x}(0),0)> 0$ and it furthermore holds that $\mathfrak{b}_l(\boldsymbol{x}(0),0)\le h_l(\boldsymbol{x}(0))-r$ if $t^*_l=0$ such that a satisfaction with a robustness of $r$ is possible. By the choice of $\gamma_{l,\infty}$ and $\mathfrak{l}_l$, it is ensured that $\mathfrak{b}_l(\boldsymbol{x}(t^\prime),t^\prime)\le h_l(\boldsymbol{x}(t^\prime))-r$ for all $t^\prime\ge t^*_l$. Hence, if now $\mathfrak{b}_l(\boldsymbol{x}(t^\prime),t^\prime)\ge 0$ for all $t^\prime\ge t^*_l$, then it follows that $ h_l(\boldsymbol{x}(t^\prime))-r\ge 0$, which implies $h_l(\boldsymbol{x}(t^\prime))\ge r$ leading to $\rho^{\phi}(\boldsymbol{x},0)\ge r$ by the choice of $t^*_l$ and $r$.

\textbf{Step 2)} For $p>1$, a more elaborate procedure is needed. Let, similarly to Step~1, $\mathfrak{b}_l(\boldsymbol{x},t):=-\gamma_l(t)+h_l(\boldsymbol{x})$ with $\gamma_l(t)$ according to \eqref{eq:t_star} and \eqref{eq:gamma}. To ensure that $\mathfrak{b}(\boldsymbol{x},t)$ is a cCBF satisfying Assumption \ref{ass3}, we pose the following assumption.
\begin{assumption}\label{ass4}
Each predicate function contained in $\phi$, denoted by $h_l(\boldsymbol{x}):\mathbb{R}^n\to\mathbb{R}$ with $l\in\{1,\hdots,p\}$, is concave.
\end{assumption}
\begin{lemma}\label{lemma:concave}
Under Assumption~\ref{ass4} and for a fixed $t'$, \begin{small}$\mathfrak{b}(\boldsymbol{x},t'):=-\frac{1}{\eta}\ln\big(\sum_{l=1}^p\mathfrak{o}_l(t')\exp(-\eta\mathfrak{b}_l(\boldsymbol{x},t'))\big)$\end{small} is concave.

\begin{proof}
Omitted due to space limitations.
\end{proof}
\end{lemma}

Compared to Step 1, it is not enough to select $\gamma_{l,0}$ as in \eqref{eq:gamma0} to ensure $\mathfrak{b}(\boldsymbol{x}(0),0)\ge 0$ due to \eqref{eq:under_approx}.  To see this, consider $\mathfrak{b}(\boldsymbol{x},t):=-\frac{1}{\eta}\ln\big(\exp(-\eta\mathfrak{b}_1(\boldsymbol{x},t))+\exp(-\eta\mathfrak{b}_2(\boldsymbol{x},t))\big)$. If $\mathfrak{b}_1(\boldsymbol{x}(0),0)> 0$ and $\mathfrak{b}_2(\boldsymbol{x}(0),0)> 0$ (which is both ensured by \eqref{eq:gamma0}), then it does not neccessarily hold that $\mathfrak{b}(\boldsymbol{x}(0),0)\ge 0$ depending on the value of $\eta$. Therefore, $\eta$ now needs to be selected sufficiently large hence increasing the accuracy of the used approximation. More importantly,  $\gamma_{l,\infty}$, which has to be selected according to  \eqref{eq:gammainfty}, and $r$ need to be selected so that for all $t\in[s_0,s_{q}]$ there exists $\boldsymbol{x}\in\mathbb{R}^n$ so that $\mathfrak{b}(\boldsymbol{x},t)\ge 0$. These objectives can be achieved by appropriately selecting $\eta$, $r$, $\gamma_{l,0}$, and $\gamma_{l, \infty}$. We formulate this parameter selection as an optimization problem that can be solved offline. Define next 
$\boldsymbol{\gamma}_{0}:=\begin{bmatrix} \gamma_{1,0} & \hdots & \gamma_{p,0}\end{bmatrix}^T$, $\boldsymbol{\gamma}_{\infty}:=\begin{bmatrix} \gamma_{1,\infty} & \hdots & \gamma_{p,\infty}\end{bmatrix}^T$, and $\boldsymbol{\mathfrak{l}}:=\begin{bmatrix} \mathfrak{l}_1 & \hdots & \mathfrak{l}_p\end{bmatrix}^T$ that contain the parameters $\gamma_{l,0}$, $\gamma_{l,\infty}$, and $\mathfrak{l}_l$ for each eventually- and always-operator encoded in $\mathfrak{b}_l(\boldsymbol{x},t)$. Define also $\boldsymbol{\xi}_1,\hdots,\boldsymbol{\xi}_q\in \mathbb{R}^n$ and let $\boldsymbol{\xi}:=\begin{bmatrix}
{\boldsymbol{\xi}_1}^T & \hdots & {\boldsymbol{\xi}_q}^T
\end{bmatrix}^T$. As argued in Section \ref{background_bf}, there also needs to exist a compact set $\mathfrak{D}'$, which is realized by including an additional barrier function $\mathfrak{b}_{p+1}(\boldsymbol{x},t)=D-\|\boldsymbol{x}\|$ for a suitably selected $D$ into $\mathfrak{b}(\boldsymbol{x},t)$. Next, select the parameters $\eta$, $r$, $D$, $\boldsymbol{\gamma}_{0}$, $\boldsymbol{\gamma}_{\infty}$, and $\boldsymbol{\mathfrak{l}}$ according to the solution of the following optimization problem
\begin{subequations}\label{eq:cbf_selection}
\begin{align}
&\underset{\eta,r,D,\boldsymbol{\gamma}_{0},\boldsymbol{\gamma}_{\infty},\boldsymbol{\mathfrak{l}},\boldsymbol{\xi}}{\operatorname{argmax}} r\label{optim_a}\\
\text{s.t.} \; &\mathfrak{b}(\boldsymbol{x}(0),0)\ge \delta \label{optim_b}\\
&\lim_{\tau\to s_j^-}\mathfrak{b}(\boldsymbol{\xi}_j,\tau)\ge \delta \;\;\; \text{for each } j\in\{1,\hdots,q \} \label{optim_c}\\
& \gamma_{l,0} \text{ as in } \eqref{eq:gamma0} \text{ for each } l\in\{1,\hdots,p+1\}\label{optim_d}\\
& \gamma_{l,\infty} \text{ as in } \eqref{eq:gammainfty} \text{ for each } l\in\{1,\hdots,p+1\}\label{optim_e}\\
& \mathfrak{l}_l \text{ as in } \eqref{eq:l} \text{ for each } l\in\{1,\hdots,p+1\}\label{optim_f}\\
& \eta> 0 \;\; \text{and} \;\; r> 0\label{optim_f}.
\end{align}
\end{subequations}
where $\delta\ge 0$. Note that $\lim_{\tau\to s_j^-}\mathfrak{b}(\boldsymbol{\xi}_j,\tau)$ can easily be evaluated since $\mathfrak{o}_l(t)$ is piecewise continuous. The optimization problem \eqref{eq:cbf_selection} is nonconvex, but can be solved offline. If maximization of $r$ is not of interest, then a feasibility program with  constraints  \eqref{optim_b}-\eqref{optim_f} can be solved instead. 
\begin{lemma}\label{lemma:cCBF}
Under Assumption \ref{ass4}, the function $\mathfrak{b}(\boldsymbol{x},t)$ obtained by the solution of \eqref{eq:cbf_selection} is a cCBF for each $[s_{j},s_{j+1})$.

\begin{proof}
Omitted due to space limitations.
\end{proof}
\end{lemma}

\begin{lemma}\label{lemma:vCBF}
Assume that \eqref{eq:cbf_selection} is solved for $\delta>0$, then $\alpha$ can be selected such that $\mathfrak{b}(\boldsymbol{x},t)$ satisfies Assumption \ref{ass3}.

\begin{proof}
\re{For each $t'\in[s_j,s_{j+1})$,  $\mathfrak{b}(\boldsymbol{x}^*_{t'},t')>\mathfrak{b}(\boldsymbol{x},t')$ for all $\boldsymbol{x}\neq\boldsymbol{x}^*_{t'}$ where $\boldsymbol{x}^{*}_{t'}:=\text{argmax}_{\boldsymbol{x}\in\mathbb{R}^n}\mathfrak{b}(\boldsymbol{x},t')$ and $\frac{\partial \mathfrak{b}(\boldsymbol{x},t')}{\partial \boldsymbol{x}}=\boldsymbol{0}$ if and only if $\boldsymbol{x}=\boldsymbol{x}^{*}_{t'}$ due Lemma \ref{lemma:concave}. Hence, $\mathfrak{b}(\boldsymbol{x}^{*}_{t'},t')\ge \delta>0$ for each $t'\in[s_0,s_{q}]$ due to \eqref{optim_b}, \eqref{optim_c}, and since $\gamma_l(t)$ is increasing so that $0<\delta\le \mathfrak{b}_l(\boldsymbol{x}^*_{t'},t')$ if $\mathfrak{o}_l(t')=1$ due to \eqref{eq:under_approx}. There also exists $\mathfrak{b}_l^\text{max}\ge 0$ such that $\mathfrak{b}_l(\boldsymbol{x}^*_{t'},t')\le \mathfrak{b}_l^\text{max}$ if $\mathfrak{o}_l(t')=1$ due to continuity of $\mathfrak{b}_l(\boldsymbol{x},t')$ on $\mathfrak{C}(t')$. Let $\mathfrak{b}^\text{max}:=\max(\mathfrak{b}_1^\text{max},\hdots,\mathfrak{b}_{p+1}^\text{max})$ and $\Delta_l:=\sup_{t\ge 0}|\frac{\partial \mathfrak{b}_l(\boldsymbol{x},t)}{\partial t}|=\mathfrak{l}_l(\gamma_{l,0}-\gamma_{l,\infty})$. Hence, we can deduce that $\frac{\partial \mathfrak{b}(\boldsymbol{x}^*_{t'},t')}{\partial t}\ge \frac{-\exp(-\eta\delta)\Delta_l}{\exp(-\eta\mathfrak{b}^\text{max})}=:\zeta$ (details omitted due to space limitations; note, however, that $\frac{\partial \mathfrak{b}_l(\boldsymbol{x},t)}{\partial t}$ is non-positive). If it is now guaranteed that $\zeta> -\alpha(\delta)$, then it holds that $\frac{\partial \mathfrak{b}(\boldsymbol{x}^*_{t'},t')}{\partial t}> -\alpha(\mathfrak{b}(\boldsymbol{x}^*_{t'},t')$ for all $t'\in[s_0,s_q]$ so that Assumption \ref{ass3} holds. By the specific choice of $\alpha(\delta)=\kappa\delta$, we can select $\kappa> \frac{-\zeta}{\delta}$ such that this is the case.}
\end{proof}
\end{lemma}

\begin{corollary}\label{corr3}
Consider the same assumptions as in Theorem \ref{theorem1}. If each $\phi_k$ additionally satisfies Assumption  \ref{ass4}, $\mathfrak{b}^k(\bar{\boldsymbol{x}}_k,t)$ is the solution of \eqref{eq:cbf_selection} for $\delta>0$, and $\alpha_k:=\kappa\delta$ with $\kappa> -\frac{\zeta}{\delta}$, then  $\rho^{\phi_k}(\bar{\boldsymbol{x}}_k,0)\ge r_k> 0$ where $r_k$ is obtained by the solution of \eqref{eq:cbf_selection} for each $k\in\{1,\hdots,K\}$. 

\begin{proof}
Follows by Theorem \ref{theorem1} and Lemmas \ref{lemma:cCBF} and \ref{lemma:vCBF}.
\end{proof}
\end{corollary}

\section{Simulations}
\label{sec:simulations}

\begin{figure*}[tbh]
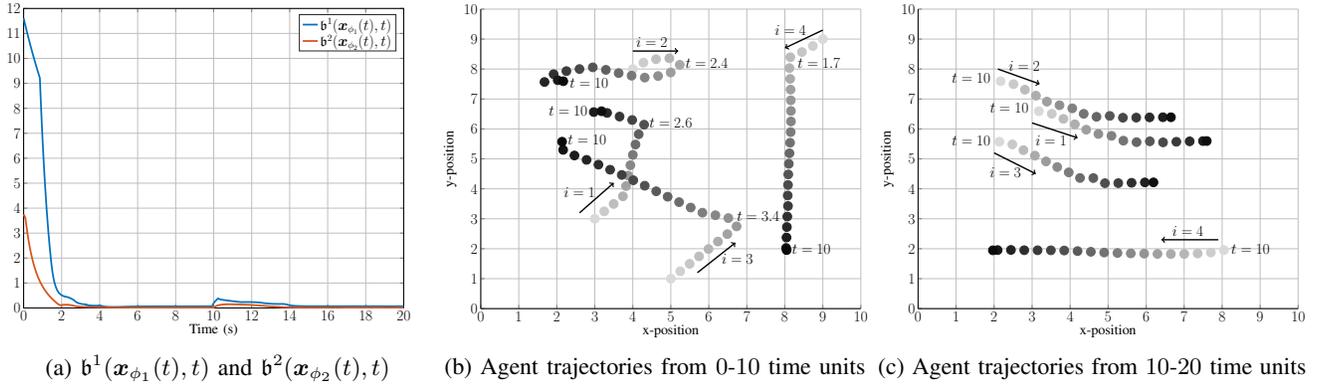

\centering
\begin{subfigure}{0.32\textwidth}
\input{figures/1}\caption{$\mathfrak{b}^1(\boldsymbol{x}_{\phi_1}(t),t)$ and $\mathfrak{b}^2(\boldsymbol{x}_{\phi_2}(t),t)$}\label{fig:1}
\end{subfigure}
\begin{subfigure}{0.32\textwidth}
\input{figures/2}\caption{Agent trajectories from $0$-$10$ time units}\label{fig:2}
\end{subfigure}
\begin{subfigure}{0.32\textwidth}
\input{figures/3}\caption{Agent trajectories from $10$-$20$ time units}\label{fig:3}
\end{subfigure}
\caption{Barrier function evolution and agent trajectories for the simulated example.}
\vspace{-10 pt}
\end{figure*}

Consider $M:=4$ agents with $\boldsymbol{x}_i:=\begin{bmatrix}x_{i,1} & x_{i,2} \end{bmatrix}^T\in\mathbb{R}^2$. The dynamics are $\dot{\boldsymbol{x}}_i=\boldsymbol{u}_i+c_i(\boldsymbol{x},t)$ where $c_i(\boldsymbol{x},t):=f_i^\text{c}(\boldsymbol{x})+\boldsymbol{w}(t)$ with $f_i^\text{c}(\boldsymbol{x}):=0.5\text{sat}_{1}(\boldsymbol{x}_4-\boldsymbol{x}_i)$ for $i\in\{1,2,3\}$ and 
$f_4^\text{c}(\boldsymbol{x}):=0.25(\text{sat}_{1}(\boldsymbol{x}_1-\boldsymbol{x}_4)+\text{sat}_{1}(\boldsymbol{x}_2-\boldsymbol{x}_4))$ are \emph{given dynamical couplings}, $\boldsymbol{w}_i\in\{\boldsymbol{w}_i\in\mathbb{R}^2|\|\boldsymbol{w}_i\|\le 0.1\}$ is noise, and where, for $\boldsymbol{\zeta}:=\begin{bmatrix}
\zeta_1 & \zeta_2
\end{bmatrix}^T\in\mathbb{R}^2$, $\text{sat}_1(\boldsymbol{\zeta}):=\begin{bmatrix}\bar{\zeta}_1 & \bar{\zeta}_2 \end{bmatrix}^T$ with $\bar{\zeta}_c=\zeta_c$ if $|\zeta_c|\le 1$, $\bar{\zeta}_c=1$ if $\zeta_c> 1$, and $\bar{\zeta}_c=-1$ if $\zeta_c<-1$ for $c\in\{1,2\}$. We use \emph{induced dynamical couplings} to avoid collisions and set $\boldsymbol{u}_i:=f_i^\text{u}(\boldsymbol{x})+\boldsymbol{v}_i$ where \re{$f_i^\text{u}(\boldsymbol{x}):=\sum_{\mathfrak{j}=1}^{3}\frac{\boldsymbol{x}_i-\boldsymbol{x}_{\mathfrak{j}}}{\|\boldsymbol{x}_i-\boldsymbol{x}_{\mathfrak{j}}\|+0.01}$ for $i\in\{1,2,3\}$ and $f_4^\text{u}(\boldsymbol{x}):=\boldsymbol{0}$. We also rewrite $(\|\boldsymbol{x}_i\|_\infty\le 1)$ as $(x_{i,1}\le 1)\wedge (-x_{i,1}\le 1)\wedge (x_{i,2}\le 1)(-x_{i,2}\le 1)$}. Let $\phi_1:=G_{[5,10]}(\|\boldsymbol{x}_1-\boldsymbol{p}_A\|_\infty\le 0.5)\wedge \big((\|\boldsymbol{x}_2-\begin{bmatrix} x_{1,1}-1 & x_{1,2}+1 \end{bmatrix}^T\|_\infty\le 0.5)\wedge (\|\boldsymbol{x}_3-\begin{bmatrix} x_{1,1}-1 & x_{1,2}-1 \end{bmatrix}^T\|_\infty\le 0.5)\big)U_{[10,20]}(\|\boldsymbol{x}_1-\boldsymbol{p}_B\|_\infty\le 0.5)$ with $\boldsymbol{p}_A:=\begin{bmatrix} 2.5 & 7 \end{bmatrix}^T$, $\boldsymbol{p}_B:=\begin{bmatrix} 8 & 6 \end{bmatrix}^T$, i.e., agent~$1$ should always be in region $\boldsymbol{p}_A$ within $5$-$10$ time units, e.g., to prepare an object for transportation, while agents $2$ and $3$ form a formation with agent $1$ from $10$ time units and until agent $1$ reaches region $\boldsymbol{p}_B$, e.g., collaborative transportation of this object. Let $\phi_2:=F_{[5,10]}(\|\boldsymbol{x}_1-\boldsymbol{p}_C\|_\infty\le 1)\wedge G_{[0,10]} (x_{4,1}\ge 8) \wedge F_{[15,20]}(\|\boldsymbol{x}_1-\boldsymbol{p}_D\|_\infty\le 1) \wedge G_{[10,20]} (x_{4,2}\le 2)$ with $\boldsymbol{p}_C:=\begin{bmatrix} 9 & 1 \end{bmatrix}^T$, $\boldsymbol{p}_D:=\begin{bmatrix} 1 & 1 \end{bmatrix}^T$, e.g., a surveillance task. We obtain $\mathfrak{b}^1(\bar{\boldsymbol{x}}_1,t)$ and $\mathfrak{b}^2(\bar{\boldsymbol{x}}_2,t)$ for $\phi_1$ and $\phi_2$, respectively; \eqref{eq:cbf_selection} has been solved as a feasibility problem with computation times of $50$ and $13$ seconds, respectively, on a two-core $1.8$ GHz CPU with 4 GB of RAM. Computing \eqref{eq:qp_conv} took, on average, $0.006$ seconds. The control barrier functions are plotted in Fig. \ref{fig:1}, while Fig. \ref{fig:2} and \ref{fig:3} show the agent trajectories from $0$-$10$ and from $10$-$20$ time units, respectively. In Fig. \ref{fig:2} agents $1$, $2$, $3$, and $4$ first approach each other due to $f_i^\text{c}(\boldsymbol{x})$. At $2.6$, $2.4$, $3.4$, and $1.7$ time units, respectively, the barrier functions force the agents to not approach each other any further and instead work towards satisfying $\phi_1$ and $\phi_2$. At $10$ time units, agent $1$ reaches region $\boldsymbol{p}_A$ while agents $2$ and $3$ form a formation. Fig. \ref{fig:3} shows how this formation is remained while agent $1$ approaches $\boldsymbol{p}_B$. Agents do not get too close to each other due to $f_i^\text{u}(\boldsymbol{x})$. It holds that $\rho^{\phi_1\wedge \phi_2}(\boldsymbol{x},0)\ge 0.005$.

\section{Conclusion}
\label{sec:conclusion}
 
We proposed decentralized control barrier functions for multi-agent systems under STL tasks. We also presented a procedure to construct control barrier functions for a particular class of STL tasks.  This last step is subject to future work to account for a more general class of STL tasks. %

\bibliographystyle{IEEEtran}
\bibliography{literature}

\begin{thebibliography}{10}
\providecommand{\url}[1]{#1}
\csname url@samestyle\endcsname
\providecommand{\newblock}{\relax}
\providecommand{\bibinfo}[2]{#2}
\providecommand{\BIBentrySTDinterwordspacing}{\spaceskip=0pt\relax}
\providecommand{\BIBentryALTinterwordstretchfactor}{4}
\providecommand{\BIBentryALTinterwordspacing}{\spaceskip=\fontdimen2\font plus
\BIBentryALTinterwordstretchfactor\fontdimen3\font minus
  \fontdimen4\font\relax}
\providecommand{\BIBforeignlanguage}[2]{{%
\expandafter\ifx\csname l@#1\endcsname\relax
\typeout{** WARNING: IEEEtran.bst: No hyphenation pattern has been}%
\typeout{** loaded for the language `#1'. Using the pattern for}%
\typeout{** the default language instead.}%
\else
\language=\csname l@#1\endcsname
\fi
#2}}
\providecommand{\BIBdecl}{\relax}
\BIBdecl

\bibitem{ren2005consensus}
W.~Ren and R.~W. Beard, ``Consensus seeking in multiagent systems under
  dynamically changing interaction topologies,'' \emph{IEEE Trans. Autom.
  Control}, vol.~50, no.~5, pp. 655--661, 2005.

\bibitem{tanner2003stable}
H.~G. Tanner, A.~Jadbabaie, and G.~J. Pappas, ``Stable flocking of mobile
  agents, part i: Fixed topology,'' in \emph{Proc. Conf. Decis. Control}, Maui,
  HI, December 2003, pp. 2010--2015.

\bibitem{zavlanos2008distributed}
M.~M. Zavlanos and G.~J. Pappas, ``Distributed connectivity control of mobile
  networks,'' \emph{IEEE Trans. Robot.}, vol.~24, no.~6, pp. 1416--1428, 2008.

\bibitem{maler2004monitoring}
O.~Maler and D.~Nickovic, ``Monitoring temporal properties of continuous
  signals,'' in \emph{Proc. Int. Conf. FORMATS FTRTFT}, Grenoble, France,
  September 2004, pp. 152--166.

\bibitem{donze2}
A.~Donz{\'e} and O.~Maler, ``Robust satisfaction of temporal logic over
  real-valued signals,'' in \emph{Proc. Int. Conf. FORMATS}, Klosterneuburg,
  Austria, September 2010, pp. 92--106.

\bibitem{raman1}
V.~\text{Raman} \emph{et al.}, ``Model predictive control with signal temporal
  logic specifications,'' in \emph{Proc. Conf. Decis. Control}, Los Angeles,
  CA, December 2014, pp. 81--87.

\bibitem{lindemann2016robust}
L.~Lindemann and D.~V. Dimarogonas, ``Robust control for signal temporal logic
  specifications using discrete average space robustness,'' \emph{Automatica},
  vol. 101, pp. 377--387, 2019.

\bibitem{liu2017distributed}
Z.~\text{Liu} \emph{et al.}, ``Distributed communication-aware motion planning
  for multi-agent systems from stl and spatel specifications,'' in \emph{Proc.
  Conf. Decis. Control}, Melbourne, Australia, December 2017, pp. 4452--4457.

\bibitem{pant2018fly}
Y.~\text{Pant} \emph{et al.}, ``Fly-by-logic: control of multi-drone fleets
  with temporal logic objectives,'' in \emph{Proc. Int. Conf. Cyber-Physical
  Syst.}, Porto, Portugal, April 2018, pp. 186--197.

\bibitem{lindemann2018decentralized}
L.~Lindemann and D.~V. Dimarogonas, ``Decentralized robust control of coupled
  multi-agent systems under local signal temporal logic tasks,'' in \emph{Proc.
  American Control Conf.}, Milwaukee, WI, June 2018, pp. 1567--1573.

\bibitem{lindemann2018control}
------, ``Control barrier functions for signal temporal logic tasks,''
  \emph{IEEE Control Syst. Lett.}, vol.~3, no.~1, pp. 96--101, 2019.

\bibitem{wieland2007constructive}
P.~Wieland and F.~Allg{\"o}wer, ``Constructive safety using control barrier
  functions,'' in \emph{Proc. IFAC Symp. Nonlin. Control Syst.}, Pretoria,
  South Africa, August 2007, pp. 462--467.

\bibitem{ames2017control}
A.~D. Ames, X.~Xu, J.~W. Grizzle, and P.~Tabuada, ``Control barrier function
  based quadratic programs for safety critical systems,'' \emph{IEEE Trans.
  Autom. Control}, vol.~62, no.~8, pp. 3861--3876, 2017.

\bibitem{wang2017safety}
L.~Wang, A.~D. Ames, and M.~Egerstedt, ``Safety barrier certificates for
  collisions-free multirobot systems,'' \emph{IEEE Trans. Robot.}, vol.~33,
  no.~3, pp. 661--674, 2017.

\bibitem{glotfelter2017nonsmooth}
P.~Glotfelter, J.~Cort{\'e}s, and M.~Egerstedt, ``Nonsmooth barrier functions
  with applications to multi-robot systems,'' \emph{IEEE Control Syst. Lett.},
  vol.~1, no.~2, pp. 310--315, 2017.

\bibitem{xu2018constrained}
X.~Xu, ``Constrained control of input--output linearizable systems using
  control sharing barrier functions,'' \emph{Automatica}, vol.~87, pp.
  195--201, 2018.

\bibitem{srinivasan2018control}
M.~Srinivasan, S.~Coogan, and M.~Egerstedt, ``Control of multi-agent systems
  with finite time control barrier certificates and temporal logic,'' in
  \emph{Proc. Conf. Decis. Control}, Miami,FL, Dec. 2018, pp. 1991--1996.

\bibitem{paden1987calculus}
B.~Paden and S.~Sastry, ``A calculus for computing filippov's differential
  inclusion with application to the variable structure control of robot
  manipulators,'' \emph{IEEE Trans. Circ. Syst.}, vol.~34, no.~1, pp. 73--82,
  1987.

\bibitem{cortes2008discontinuous}
J.~Cortes, ``Discontinuous dynamical systems,'' \emph{IEEE Control Syst. Mag.},
  vol.~28, no.~3, 2008.

\bibitem{shevitz1994lyapunov}
D.~Shevitz and B.~Paden, ``Lyapunov stability theory of nonsmooth systems,''
  \emph{IEEE Trans. Autom. Control}, vol.~39, no.~9, pp. 1910--1914, 1994.

\bibitem{filippov2013differential}
A.~F. Filippov, \emph{Differential equations with discontinuous righthand
  sides: control systems}.\hskip 1em plus 0.5em minus 0.4em\relax Springer
  Science \& Business Media, 2013, vol.~18.

\end{thebibliography}

\addtolength{\textheight}{-12cm}   

\end{document}